\documentclass[twocolumn,aps,floats,floatfix,prl,nofootinbib,superscriptaddress,tightenlines]{revtex4-1}

\usepackage{graphicx}
\usepackage{bm}
\usepackage{epsfig}
\usepackage{pstricks}
\usepackage{amsmath}
\usepackage{hyperref}
\usepackage{color}
\usepackage[normalem]{ulem}
\input colordvi

\newcommand{\beq}{\begin{equation}}
\newcommand{\eeq}{\end{equation}}
\newcommand{\bea}{\begin{eqnarray}}
\newcommand{\eea}{\end{eqnarray}}
\newcommand{\OMIT}[1]{}

\begin{document}

\title{Black hole mass dynamics and  renormalization group evolution}

\def\addCMU{Department of Physics, Carnegie Mellon University, Pittsburgh PA  15213, USA}
\author{Walter D. Goldberger}\affiliation{Department  of Physics, Yale University, New Haven CT 06520, USA}
\author{Andreas Ross}
\author{Ira Z. Rothstein}
\affiliation{\addCMU}

\begin{abstract}

We examine the real-time dynamics of a system of one or more black holes interacting with long wavelength gravitational fields.   We find that the (classical) renormalizability of the effective field theory that describes this system necessitates the introduction of a time dependent mass counterterm, and consequently the mass parameter must be promoted to a dynamical degree of freedom.     To track the time evolution of this dynamical mass, we compute the expectation value of the energy-momentum tensor within the in-in formalism, and fix the time dependence by imposing energy-momentum conservation.   Mass renormalization induces logarithmic ultraviolet divergences at quadratic order in the gravitational coupling, leading to a new time-dependent renormalization group (RG) equation for the mass parameter.  We solve this RG  equation and use the result to predict heretofore unknown high order logarithms in the energy distribution of {gravitational radiation emitted} from the system.
\end{abstract} 

\maketitle 

\section{Introduction}

Understanding the dynamics of a system of gravitationally bound classical or quantum black holes is a problem of both theoretical and observational interest.     To have any hope of finding analytic solutions to this non-linear problem necessitates some sort of approximation scheme.   An obvious avenue for finding such solutions lies in the limit where the typical frequency $\omega$ of probes of this system (e.g. outgoing radiation) is much {smaller} than the inverse of the {size of the system and its} gravitational radius $r_s$.   

 As we discussed in previous work~\cite{NRGR,Goldberger:2005cd}, in this limit the dynamics is summarized by an effective worldline Lagrangian coupled to bulk gravity, describing the evolution of a suitable center of mass coordinate $x^\mu(\lambda)$.    The worldline theory is of the form
\begin{align}
\label{eq:S}
 S = & - \! \int d\tau(\lambda)  M(\lambda) + \frac{1}{2} \int d\tau(\lambda) I^{ij}(\lambda) E_{ij}(\lambda) + \dots.
\end{align}
 In this equation $\lambda$ is an affine parameter, $\tau$ is proper time measured using the bulk metric $g_{\mu\nu}$ and $E_{ij}$ is a projection of the bulk Weyl tensor onto its electric-type parity components, $E_{\mu\nu}(\lambda)=C_{\mu\alpha\nu\beta} (x(\lambda)){\dot x}^\alpha {\dot x}^\beta$.   The objects $M(\lambda)$, $I_{ij}(\lambda)$ describe the first two dynamical $\ell=0,2$ $SO(3)$ multipole moments of the system.    Ellipses denote higher order moments which do not play a role in our present discussion. 
   See~\cite{Goldberger:2005cd,Goldberger:2009qd,Ross:2012fc} for a more detailed discussion of this Lagrangian.

The Lagrangian in Eq.~(\ref{eq:S}) is suitable for describing either the quantum or classical dynamics of an ensemble of black holes in the long wavelength limit.    In the quantum theory, the moments are operators acting on the Hilbert space of black hole microstates.    We showed in ref.~\cite{Goldberger:2005cd} that even without complete knowledge of the microscopic theory, it is possible to obtain some information on the various correlators of these worldline operators through a matching procedure to bulk graviton scattering and absorption processes, in a way that is similar to the CFT description of bulk black holes in AdS/CFT.   

Alternatively, Eq.~(\ref{eq:S}) may be viewed classically, in which case the moments are fixed by first treating the constituents as point particles and then integrating out all of the modes whose wavelength is of order the orbital separation to generate a theory of a single composite object.    The dynamics of the moments are dictated by solving the equations of motion for the constituents. Consistency requires that this evolution should account not only for the interparticle potentials but also for  the effects of self-forces which encapsulate both conservative (backscattering of radiation from the system)  as well as non-conservative (radiation recoil) effects.   An effective field theory description of this procedure in the non-relativistic (or ``post-Newtonian") limit  was first formulated in~\cite{NRGR}, and extended in~\cite{Goldberger:2009qd} to include higher order radiative corrections {and in \cite{Galley:2009px} to account for radiation reaction}.

Our focus in this paper is on the $\ell=0$ mass mode $M$ and its time evolution due to  gravitational radiation.  Although this time dependence may be neglected to an excellent approximation at low orders in the case of nearly-adiabatic, post-Newtonian systems, at higher orders the time dependence plays a role in the dynamics and must be consistently accounted for.  

We find that time dependence is necessary from the point of view of the renormalizability of the theory.   Even if the mass parameter is made constant at some renormalization scale, renormalization group flow at the classical level induces time dependence.   The fact that the  mass of a radiating  system changes with time
implies that our mass is not the ADM mass. In fact, as we shall see, the boundary value of the mass, in terms of the
renormalization group scale, will be the Bondi mass.
  We also obtain a time evolution equation for the renormalized mass parameter.    In order to do this, we impose conservation of a suitably defined energy-momentum {(pseudo-)}tensor operator for the composite system of gravitating sources, whose expectation value is computed using the closed time path formalism familiar from non-equilibrium quantum field theory and statistical mechanics {\cite{inin}}, here adapted to the classical system {(see \cite{Galley:2012hx} for a purely classical formulation)}.   Finally, as a byproduct of our analysis, we use the solution to the coupled renormalization group equations for $M$ and  $I_{ij}$ to predict the pattern of logarithms that appear in the distribution of low-frequency $\ell=2$ gravitons radiated from the bound system.

\section{Renormalization and time dependence}

In what follows, we work in the classical limit, taking the moments in Eq.~(\ref{eq:S}) as c-number sources.    Gravitational dynamics is taken to follow from the Einstein-Hilbert action, $S_{EH}=-{1\over 16 \pi G} \int d^4 x \sqrt{g} R$.    Our goal is to set up a causal time evolution equation for the point sources in Eq.~(\ref{eq:S}) as they radiate gravitons.   The correct language for this sort of dissipative, causal problem is the in-in formalism~\cite{inin}.   We expand the metric $g_{\mu\nu} =  {\bar g}_{\mu\nu} + h_{\mu\nu}$ into a weak background ${\bar g}_{\mu\nu}=\eta_{\mu\nu}+ {\bar h}_{\mu\nu}$ and a fluctuating field $h_{\mu\nu}$ and integrate out the fluctuations according to the in-in prescription.   This yields an effective action
\begin{align}
e^{i\Gamma[{\bar h}_+,{\bar h}_-,I,M]} = \int D h^+(x) Dh^-(x) e^{iS[g^+,M,I]-iS[g^-,M,I]},
\end{align}
where we double the number of degrees of freedon and impose closed-time path boundary conditions on the fluctuating modes.    In this paper we treat the multipole {moments} as classical, so it is not necessary to double them {for our purposes}.    The main object of interest is the real-time expectation value of the effective energy momentum tensor, which is 
\begin{equation}
\langle in|T_{\mu\nu}(x)| in\rangle = - \left.{2\over \sqrt{g^+}} {\delta \Gamma\over \delta {\bar g}_+^{\mu\nu}(x)}\right|_{{\bar h}^+={\bar h}^-=0}.
\end{equation}
As a consequence of Ward identities, it obeys $\partial_\mu \langle T^{\mu\nu}(x)\rangle=0$, which in fact fully determines the evolution of the moments.

\begin{figure}[!t]
\centerline{{\includegraphics[scale=0.36]{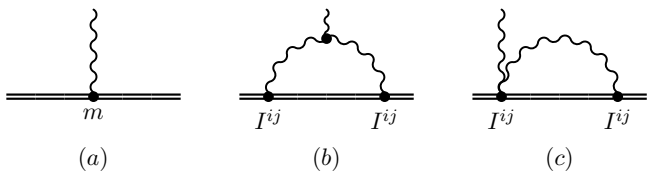}}}
\caption[1]{Leading diagram topologies for the computation of $\left<T^{\mu\nu}\right>$.} \label{fig:LO}
\end{figure}

Diagrammatically, $\langle T_{\mu\nu}\rangle$ consists of graphs with one external graviton ${\bar h}^+_{\mu\nu}$ (or equivalently  ${\bar h}^-_{\mu\nu}$).   To first non-trivial order in the long wavelength approximation, the relevant diagram topologies are given in Fig.~\ref{fig:LO}.   Also at this order  in the expansion,  we may set the spatial momentum of the external graviton to zero in the calculation of  $\langle T_{\mu\nu}\rangle$.     We also assume that the center of mass worldline is at rest at the origin $x^\mu(\lambda)=({t},0,0,0)$.    Then the diagrams in Fig.~\ref{fig:LO} all give $\langle T^{0i}\rangle$ of order $\nabla \delta^3(\mathbf x)$, and from Fig.~\ref{fig:LO}(a),
\begin{equation}
\label{eq:LO}
\langle T^{00}(x)\rangle = M({t}) \delta^3({\bf x}),   
\end{equation}
so that at this order the conservation law $\partial_0\langle T^{00}(x)\rangle =0$ simply yields ${\dot M}=0$.

The diagrams in Fig.~\ref{fig:LO}(b), (c) have a more interesting structure.    Although these diagrams do not contain logarithmic divergences in $d=4$ spacetime dimensions, the computation of these terms does involve the one-loop integral
\begin{equation}
\int {d^{d-1} {\bf q}\over (2\pi)^{d-1}}  {1\over {\bf q}^2 + \omega^2} = {\Gamma(3/2-d/2)\over (4\pi)^{(d-1)/2}} \omega^{d-3},
\end{equation}
which is linearly ultraviolet divergent in $d=4$.   While such power divergences are set to zero within dimensional regularization in the $\overline{MS}$ scheme, other schemes which do not automatically set such divergences to zero would require the introduction of a time-dependent counterterm for the mass mode.    Thus even if the time dependence of $M$ is set to zero in the bare Lagrangian, renormalizability of the theory requires the mass to become dynamical.  

In the computation of the graphs in Fig.~\ref{fig:LO}(b), (c), it is sufficient to set ${\bar h}^-_{\mu\nu}=0,$ in which case one finds that the various propagators in the internal lines combine into  retarded graviton propagators.    Using the Feynman rules and the gauge-fixing choice in Ref.~\cite{NRGR}, one finds the $x$-space result
\begin{align}
\langle T^{00}\rangle_{(b)} & = \frac{G}{30} \, \delta^3(\mathbf x) \biggl(7 \hspace*{1pt} I^{(5)}_{ij}(t) I_{ij}(t) - 5 I^{(4)}_{ij}(t) I^{(1)}_{ij}(t) \biggr. \notag \\ & \biggl. \quad \quad \quad \quad \quad \quad   - 6 \int_{- \infty}^{t} dt' I^{(6)}_{ij}(t') I_{ij}(t')\biggr), \\
\langle T^{00}\rangle_{(c)} & = \frac{G}{30} \, \delta^3(\mathbf x) \left(- \hspace*{1pt} I^{(5)}_{ij}(t) I_{ij}(t) + 5 I^{(4)}_{ij}(t) I^{(1)}_{ij}(t) \right) \, .
\end{align}
Including the contribution of Eq.~(\ref{eq:LO}) this  gives the  time dependence of the $l=0$ mode  
\footnote{For related results in a different formalism see \cite{Blanchet1,Blanchet2}.}, 
\begin{align}
 \langle T^{00}(x)\rangle & = \delta^3(\mathbf x) \left[M + \frac{G}{5}  \int_{- \infty}^{t} dt' I^{(5)}_{ij}(t') I_{ij}^{(1)}(t') \right]\ \label{eq_total_OG}.
\end{align}
The in-in expectation value is causal, depending only on the past history of the system.    Energy-momentum conservation at this order in the multipole expansion imposes non-trivial time evolution for the mass mode,
\begin{align}
\label{eq:Mdot}
 \dot M(t) & = - \frac{G}{5} I^{(5)}_{ij}(t) I^{(1)}_{ij}(t) \, .
\end{align}
As a check of our result, note that upon time-averaging both sides of this equation, we recover the textbook quadrupole radiation formula $\langle {\dot M}\rangle = -{G\over 5} \langle  I^{(3)}_{ij} I^{(3)}_{ij} \rangle,$ reflecting the fact that the time dependence of the mass accounts for energy loss into gravitational radiation.

\section{Renormalization group flows}

\begin{figure}[!t]
\centerline{{\includegraphics[scale=0.36]{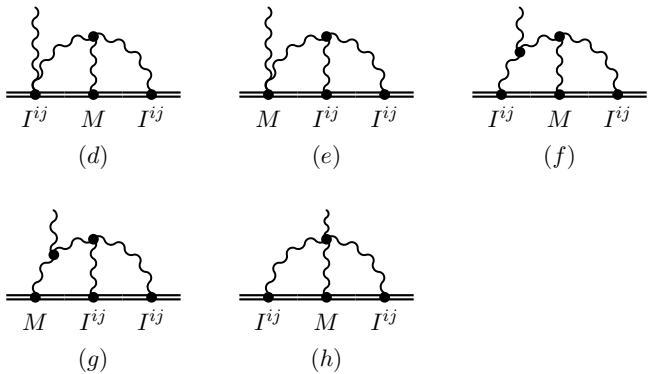}}}
\caption[1]{Leading diagram topologies which yield logarithmically UV divergent contributions to $\left<T^{\mu\nu}\right>$.} \label{fig:UV}
\vskip 0.5cm
\end{figure}

Ultraviolet divergences in four-dimensional gravity first appear in the order $G^2$ contributions to  $\langle T^{\mu\nu}\rangle$.    The relevant diagrams are given in Fig.~\ref{fig:UV}.   We focus here on obtaining renormalization group equations, and therefore we only retain the parts of the diagrams that are singular as $d\rightarrow 4$.    Because the divergences of in-in correlators are the same as in the time-ordered scheme, it is simpler to employ  time-ordered Feynman rules.     Given the results of the previous section, the time dependence of $M(t)$ would only yield contributions to the RG equation at order $G^3$, and we may safely treat $M$ as a constant in the calculation of the diagrams in Fig.~(\ref{fig:UV}).    Thus up to higher order corrections, we may replace $M(t)$ by its time average ${\bar M}=\int_{-T}^T M(t) dt/(2T)$  in the vertices.

Working in momentum space with external graviton momentum $k^\mu=(\omega,0,0,0)$, we find that the divergent parts  of the diagrams in Fig.~\ref{fig:UV} can be written as
\begin{eqnarray}
\nonumber
\mbox{Fig.~\ref{fig:UV}} &=&{i\pi^2 \over 10} G^2 {\bar M} {\cal I} \int {d\omega_1\over 2\pi} {d\omega_2\over 2\pi} I_{ij}(\omega_1) I_{ij}(\omega_2) \\
& &  \times (2\pi) \delta(\omega-\omega_1-\omega_2) P(\omega_1,\omega_2) +\cdots,
\end{eqnarray}
where we omit finite contributions.    The function $P(\omega_1,\omega_2)$ is a symmetric polynomial in its two arguments and ${\cal I}$ is the integral
\begin{eqnarray}
\nonumber
{\cal I}  &=& (\mu^2)^{4-d} \int {d^{d-1} {\bf q}\over (2\pi)^{d-1}} {d^{d-1} {\bf p}\over (2\pi)^{d-1}}    {1\over {\bf q}^2\, {\bf p}^2\, ({\bf q}+{\bf p})^2}\\
&=& -{1\over 32 \pi^2} \left[{1\over d-4} - \ln \mu^2+\cdots\right],
\end{eqnarray}
with $\mu$ an arbitrary subtraction scale.  We find for each diagram
\begin{eqnarray}
P(\omega_1,\omega_2)_{(d)} &=& 32 (\omega_1^5 \omega_2 +\omega_1 \omega_2^5),\\
P(\omega_1,\omega_2)_{(e)} &=& -{32}  \omega_1^2\omega_2^2  \left(\omega_1^2+\omega_1 \omega_2+  \omega_2^2\right),\\
\nonumber
P(\omega_1,\omega_2)_{(f)} &=&\omega_1^6 + \omega_2^6 -8 (\omega_1^5 \omega_2 + \omega_1 \omega_2^5)\\
& & - 49 (\omega_1^2 \omega_2^4 + \omega_1^4 \omega_2^4)  + 48 \omega_1^3 \omega_2^3,\\
\nonumber
P(\omega_1,\omega_2)_{(g)} &=& -{3\over 2}(\omega_1^6 + \omega_2^6) -{13\over 3} (\omega_1^5 \omega_2 + \omega_1 \omega_2^5)\\
& &  + {83\over 2}  (\omega_1^4 \omega_2^2+ \omega_1^2 \omega_2^4) + {74\over 3} \omega_1^3 \omega_2^3,\\
\nonumber
P(\omega_1,\omega_2)_{(h)} &=& {1\over 6}  (\omega_1+\omega_2)^2   \left[ -94 \omega_1^2 \omega_2^2  + 3 (\omega_1^4 + \omega_2^4)\right. \\
& & \left. + 68 (\omega_1^3 \omega_2 +\omega_1 \omega_2^3)\right],
\end{eqnarray}
giving the total result
\begin{eqnarray}
\nonumber
P(\omega_1,\omega_2) = 32\left[(\omega_1^5 \omega_2 +\omega_1 \omega_2^5) + \omega_1^3 \omega_2^3 - (\omega_1^4 \omega_2^2+\omega_1^2 \omega_2^4)\right].\\
\end{eqnarray}

The logarithmic divergence in the Feynman diagrams of Fig.~\ref{fig:UV} can be absorbed into the mass counterterm $-{i\over 2} M(\omega)$ of Fig.~\ref{fig:LO}(a), yielding a finite result for $\langle T^{00}\rangle$ at this order and a correponding renormalization group equation
\begin{eqnarray}
\nonumber
\mu {d\over d\mu} M(\omega,\mu) &=& {G^2 {\bar M}\over 80} \int {d\omega_1\over 2\pi} {d\omega_2\over 2\pi} I_{ij}(\omega_1) I_{ij}(\omega_2) \\
& &  \times (2\pi) \delta(\omega-\omega_1-\omega_2) P(\omega_1,\omega_2) ,
\end{eqnarray}
or equivalently, the local $x$-space expression
\begin{align}
 \mu {d\over d\mu} M(t,\mu) & = - \frac{2 G^2 {\bar M} }{5}\left(2 I^{(5)}_{ij} I^{(1)}_{ij} - 2 I^{(4)}_{ij} I^{(2)}_{ij} + I^{(3)}_{ij} I^{(3)}_{ij}\right)(t)\, . \label{eq:RGM}
\end{align}

In addition to mass renormalization, the theory defined by Eq.~(\ref{eq:S}) also induces logarithmic renormalization of the quadrupole moment.   This effect can be traced to singularities in graviton scattering off the gravitational field of the system, in particular from the $(G M/r)^2$ general relativistic correction to the potential.   The result, first written in~\cite{Goldberger:2009qd}, is:
\begin{align}
\label{eq:BQ}
 \mu \frac{d I_{ij}}{d \mu}(\omega, \mu) & = - \frac{214}{105} (G {\bar M} \omega)^2 I_{ij}(\omega, \mu)\, .
\end{align}

\begin{figure}[!t]
\centerline{{\includegraphics[scale=0.6]{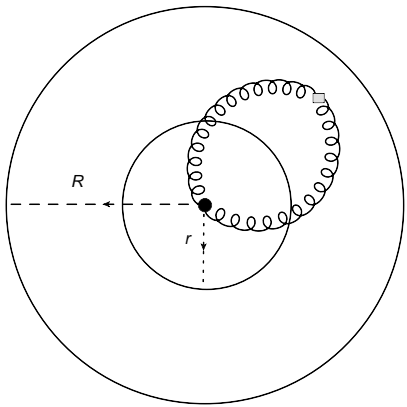}}}
\caption[1]{An on shell graviton is emitted and scattered back ( denoted by the small box)  at a distance $\rho$. Observers at different distance
would not agree on the value of the mass.} \label{Bondi}
\vskip 0.5cm
\end{figure}

There is a simple  interpretation of the mass  RG flow. The mass as parameter as defined in Eq. (\ref{eq:S}) is the energy
of the {\it conservative} system not including energy in gravitational radiation, and it is not constant in time as we showed explicitly in the last section. At the order we are working it is clear that our mass definition is the
Bondi mass, which gets no contribution from the radiation at infinity.
That this mass is scale dependent can be roughly seen by noting that the logs in the mass arise as a consequence of radiation backscattering off of the static background.
In Fig. \ref{Bondi} we illustrate such a physical process. If the graviton scatters at a radius $\rho$ then an
observer at $r<\rho$ would see radiation ``at infinity'', whereas an observer at $R>\rho$ would include
the energy of the backscattered graviton in the definition of the mass. Thus as observers reduce their distance
to the center of the black hole system the mass parameter will diminish. In this sense the mass parameter is
``asymptotically free".

%

\section{Applications}

For a gravitationally bound system of typical size ${\cal R}$ we define a velocity parameter by $v\equiv{\cal R} {\omega}$.   For low frequency radiation, with $r_s\omega \leq v\ll 1$ {where $r_s= 2 G {\bar M}$ is the gravitational radius},  the RG running of $I_{ij}(\omega,\mu)$ is an ${\cal O}[{(r_s\omega)^2}]$ effect.  On the other hand,  the running of the mass mode is  suppressed by $v^4$ relative to this effect.    Thus, to capture the leading logarithms of the form $v^4(r_s\omega)^{2n}\ln^n v$ in the mass, we may  solve the coupled RG equations by first inserting the solution to Eq.~(\ref{eq:BQ}), assuming that ${\bar M}$ is  $\mu$-independent \cite{Goldberger:2009qd},
\begin{equation}
\label{eq:Rsoln}
I_{ij}(\omega,\mu) = \left({\mu\over\mu_0}\right)^{\beta_I (G {\bar M_0}\omega)^2} I_{ij}(\omega,\mu_0),
\end{equation}
where $\beta_I= - 214/105$ the beta function coefficient in Eq.~(\ref{eq:BQ}) and ${\bar M}_0$ is the time-averaged mass, renormalized at the scale $\mu_0$.   By time averaging Eq.~(\ref{eq:RGM}), we find  the running mass ${\bar M}(\mu)$ evolves according to the equation
\begin{equation}
\mu {d\over d\mu}  \ln {\bar M} = - 2 G^2 \langle I_{ij}^{(3)} I_{ij}^{(3)} \rangle_\mu,
\end{equation}
where $\langle I_{ij}^{(3)} I_{ij}^{(3)} \rangle_\mu$ is renormalized at the scale $\mu$.   The solution is
\begin{eqnarray}
{{\bar M}(\mu)\over {\bar M}_0} &=& \exp\left[{{\langle I_{ij}^{(2)} I_{ij}^{(2)} \rangle_{\mu_0}}- \langle I_{ij}^{(2)} I_{ij}^{(2)} \rangle_\mu\over  \beta_I {\bar M}^2_0}\right], \label{eq_solRGM}
\end{eqnarray}
where in terms of time averages at a high frequency scale $\mu_0$,
\begin{eqnarray}
\nonumber
\langle I_{ij}^{(2)} I_{ij}^{(2)}\rangle_\mu &=& \sum_{n=0}^\infty {( \beta_I r_s^2)^n\over {2^n} \, n!}\langle I_{ij}^{(n+2)} I_{ij}^{(n+2)}\rangle_{\mu_0} \ln^n {\mu\over\mu_0},\\
\end{eqnarray}
with  $r_s = 2 G {\bar M}_0$.  Taking $\mu/\mu_0\sim {\cal R} \omega =v$, the terms in the series are of  the form  $(r_s\omega)^{2n}\ln^n v$.   The first few leading logs in the running mass are
\begin{eqnarray}
\nonumber
{{\bar M}(\mu)\over {\bar M}_0} &=& 1 -  \frac{1}{2}{\langle I_{ij}^{(3)} I_{ij}^{(3)}\rangle_0\over {\bar M}^2_0} r_s^2 \ln v +{107\over 420} {\langle I_{ij}^{(4)} I_{ij}^{(4)}\rangle_0\over {\bar M}^2_0} r^4_s \ln^2 v \\
& & {} -{11449\over 132300} {\langle I_{ij}^{(5)} I_{ij}^{(5)}\rangle_0\over {\bar M}^2_0} r^6_s \ln^3 v + \cdots, \label{eq_logsM}
\end{eqnarray}
and the term linear in the logarithm agrees with the result of~\cite{Blanchet:2010zd} obtained by different methods.

This result can be used to determine logarithmic terms in the graviton emission amplitude from the system.      In particular, the single ($\ell=2$) emission amplitude to order $r_s\omega$ is given by (see e.g.~\cite{Goldberger:2009qd})
\begin{equation}
\left| {\cal A}(\omega)\over {\cal A}_0(\omega)\right|^2 = 1 + 2\pi G {\bar M} \omega,
\end{equation}
where the leading order amplitude for polarized gravitons is ${\cal A}_0(\omega) = {i\over 4 m_{Pl}} \epsilon^{*}_{ij} I_{ij}(\omega, \mu),$ with $m^{-2}_{Pl} = 32 \pi G$.   Using renormalization group invariance, we may evaluate the running quantities $I_{ij}(\omega)$ and ${\bar M}$ at a scale $\mu\sim \omega$ in order to minimize possible large logarithms in higher order contributions to $\left| {\cal A}(\omega)\right|^2$.    From the solution to the RG equation for the mass of Eq. (\ref{eq_solRGM}), we obtain
\begin{align}
\left| {\cal A}(\omega)\over {\cal A}_0(\omega)\right|^2 & = 1 + \pi r_s \omega - \frac{\pi}{2}{\langle I_{ij}^{(3)} I_{ij}^{(3)}\rangle_0\over {\bar M}^2_0} r_s^3  \omega \ln v \notag \\
 & +  \frac{107 \pi}{420} {\langle I_{ij}^{(4)} I_{ij}^{(4)}\rangle_0\over {\bar M}^2_0} r^5_s  \omega \ln^2 v+ \cdots \, .
\end{align}
The logarithmic corrections scale as $\pi v^4 (r_s \omega)^{2n+1} \ln^n v$, and we note that at this order, further logarithmic terms may arise from the subleading renormalization of the quadrupole moment.

Unlike the first order contributions, the logarithmically divergent part of $\left<T^{00}\right>$ at order $G^2$ is a conservative effect.  It therefore makes sense to include it in the definition of a conserved energy $E$.  For post-Newtonian inspiralling binary systems with quasi-circular orbits, the conserved energy is commonly expressed as a function of the orbital frequency $\Omega$.  Logarithms in $E(\Omega)$ arise first at fourth post-Newtonian order \cite{Blanchet:2010zd}. There is a  direct contributions  from the energy we computed here as well 
as an indirect contribution due to the the use of the equations of motion in deriving the relation between orbital  radius and frequency.  The equations of motion may be inferred from energy conservation, by essentially reversing the arguments in \cite{Blanchet:2010zd}. We find the leading contribution linear in  logarithms to be
\begin{align}
 E(\Omega) = - \frac{\mu}{2} \, \frac{448}{15}  \nu x^5 \ln x + \dots \, ,
\end{align}
where $\mu$ is the reduced mass, $\nu = \mu/\bar M_0$ and $x = (G \bar M_0 \Omega)^{2/3}$.  This is in agreement with the 4PN logarithm computed in \cite{Blanchet:2010zd}.  We may also use our result for the running mass in Eq. (\ref{eq_solRGM}) or (\ref{eq_logsM}) to extract higher powers of the leading logarithms in $E(\Omega)$ by resumming the leading logarithms to all orders \cite{eomegaresummed}, which includes the $\ln^n x$ terms at $(4+3n)$ post-Newtonian order  including their numerical coefficient.

%
%

\section{Conclusions}

In this paper, we have set up a formalism for determining the real-time evolution of the dynamical moments that describe a system of gravitationally bound black holes.    In order to account for the dissipation of energy due to radiation in a way that respects causality, the in-in formulation of quantum field theory must be employed.   We find that the time evolution of the $\ell=0$ mass mode is inextricably linked to the renormalization of the effective theory.    At second order in the gravitational coupling $G$ one finds logarithmic ultraviolet divergences even in the classical theory, which induce non-trivial RG flows.     The RG equation for the mass mode, together with  previously obtained results for the running of the $\ell=2$ moment, resums the leading
logarithms in the conservative energy of the form $v^4(r_s \omega)^{2n} \ln ^n v.$

The results of this paper can be extended in several of directions.  It is clear that the methods introduced here can also be used to set up evolutions equations for the higher moments (the center of mass momentum and angular momentum as well as higher moments) that take into account radiative losses.   To do so would require computing the similar diagrams to those in Fig.~\ref{fig:LO}, but with non-zero spatial momentum.

Another direction of research  would be to consider the evolution of quantum black holes within this formalism, and the consequences of the RG flows for (e.g. Hawking) radiative processes.   In that case, we expect that at leading order in the multipole expansion, the  corresponding evolution equation for the mass operator will take the form of a relation between its expectation value and certain two-point correlators of the quadrupole moments, some of which have already been obtained in~\cite{Goldberger:2005cd} in the low-frequency limit.

 This work is supported by DOE grant DE-FG-02-92ER40704 (WG) and by NASA grant 22645.1.1110173 (AR, IZR).

\end{document}